\documentclass[
 reprint,superscriptaddress,
 amsmath,amssymb,
 aps,floatfix
]{revtex4-1}

\RequirePackage[utf8]{inputenc}
\RequirePackage[english]{babel}
\RequirePackage{amsmath,amsfonts,amssymb}
\RequirePackage{mathpazo} % Widely available alternative to Garamond
\RequirePackage[scaled]{helvet}
\RequirePackage[T1]{fontenc}
\RequirePackage{url}
\RequirePackage[colorlinks=true, allcolors=blue]{hyperref}
\RequirePackage{microtype}
\RequirePackage{braket}

\RequirePackage{graphicx,xcolor}
\RequirePackage{colortbl}
\RequirePackage{booktabs}
\RequirePackage{algorithm}
\RequirePackage[noend]{algpseudocode}
\RequirePackage{changepage}
\RequirePackage{mathtools}
\RequirePackage[left=48pt,%
                right=42pt,%
                top=46pt,%
                bottom=60pt,%
                headheight=15pt,%
                headsep=10pt,%
                letterpaper,twoside]{geometry}%
\RequirePackage[labelfont={bf,sf},%
                labelsep=period,%
                figurename=Fig.,%
                singlelinecheck=off,%
                justification=RaggedRight]{caption}
\begin{document}

\preprint{APS/123-QED}

\title{Generation of discord through a remote joint continuous variable measurement}

\date{\today}% It is always \today, today,

\author{E.~Zalys-Geller}
\affiliation{Department of Applied Physics, Yale University, New Haven, Connecticut 06520, USA}
\author{A.~Narla}
\affiliation{Department of Applied Physics, Yale University, New Haven, Connecticut 06520, USA}
\author{S.~Shankar}
\affiliation{Department of Applied Physics, Yale University, New Haven, Connecticut 06520, USA}
\author{M.~Hatridge}
\affiliation{Department of Applied Physics, Yale University, New Haven, Connecticut 06520, USA}
\affiliation{Department of Physics and Astronomy, University of Pittsburgh, Pittsburgh, Pennsylvania 15260, USA}
\author{M.~P.~Silveri}
\affiliation{Department of Applied Physics, Yale University, New Haven, Connecticut 06520, USA}
\affiliation{Research Unit of Nano and Molecular Systems, University of Oulu, 90014 Oulu, Finland}
\author{K.~Sliwa}
\affiliation{Department of Applied Physics, Yale University, New Haven, Connecticut 06520, USA}
\author{Z.~Leghtas}
\affiliation{Department of Applied Physics, Yale University, New Haven, Connecticut 06520, USA}
\affiliation{Centre Automatique et Systèmes, Mines ParisTech, PSL Research University, 60 Boulevard Saint-Michel, 75006 Paris, France}
\author{M.~H.~Devoret}
\affiliation{Department of Applied Physics, Yale University, New Haven, Connecticut 06520, USA}

\begin{abstract}

In quantum mechanics, continuously measuring an observable steers the system into one eigenstate of that observable. This property has interesting and useful consequences when the observable is a joint property of two remotely separated qubits. In particular, if the measurement of the two-qubit joint observable is performed in a way that is blind to single-qubit information, quantum back-action generates correlation of the discord type even if the measurement is weak and inefficient. We demonstrate the ability to generate these quantum correlations in a circuit-QED setup by performing a weak joint readout of two remote, non-interacting, superconducting transmon qubits using the two non-degenerate modes of a Josephson Parametric Converter (JPC). Single-qubit information is erased from the output in the limit of large gain and with properly tailored cavity drive pulses. Our results of the measurement of discord are in quantitative agreement with theoretical predictions, and demonstrate the utility of the JPC as a which-qubit information eraser.

\end{abstract}

\maketitle

\section{Introduction}

	Quantum measurement is traditionally treated in the instantaneously projective limit where the observation fully collapses the system into one of the eigenstates of the associated observable. Recent theoretical developments as well as experiments with well controlled quantum systems suggest that measurement should be instead understood as a progressive exchange of information between the system and an environment containing a pointer variable, or meter~\cite{Carmichael1993,Haroche2006,Wiseman,Brune1996,Guerlin2007,Katz2008,Hatridge2013,Groen2013,Murch2013,DeLange2014,Riste2013,Roch2014,Weber2014,Tan2015,Silveri2016,Campagne-Ibarcq2016}. Any information gained by the meter steers the system towards one of the eigenstates of the measurement observable. The dynamics of the latter process have been observed experimentally in the field of cavity-QED~\cite{Brune1996,Haroche2006,Guerlin2007} and circuit-QED~\cite{Katz2008,Hatridge2013,Groen2013,Murch2013,DeLange2014}
    
    Our understanding of this back-action can naturally be extended to a two qubit measurement where, for example, a joint measurement can project a two-qubit system into an entangled state~\cite{Riste2013,Roch2014,Silveri2016}. Thus, the back-action of such a joint measurement is an extremely powerful tool since it may generate non-trivial two-qubit correlations even between qubits that are remote and and have no possibility to directly interact. One observable capable of generating quantum correlations between two qubits is $IZ+ZI$ (where $Z$ and $I$ refer to the Pauli $\sigma_z$ and identity operators respectively, and the lexicographic position indexes to the corresponding subsystem). The apparatus performing this measurement learns the total number of excitations in the two-qubit system while being blind to information specific to either one of the two qubits. When this observable is measured on two qubits that are initialized in $\ket{++}$ (where $\ket{\pm}$ is the eigenstate of the operator $X$ with the eigenvalues $\pm1$), the measurement erases single-qubit phase information and generates non-trivial two-qubit phase correlations. However, observing this back-action is difficult in practice as the correlations are reduced by noise arising from coupling between each qubit and its local environment. Such noise in the joint measurement process tends to reduce two-qubit correlations below the threshold for provable entanglement.
    
Nevertheless, the back-action of a joint measurement can manifest in other uniquely quantum signatures due to the properties of quantum mechanics. A surprising characteristic of quantum measurement is that a measurement which has \textit{any} blindness whatsoever to local properties can result in a back-action which cannot be reproduced by local measurements alone. Even with a large amount of local noise, the process of measuring $IZ+ZI$ is physically distinct from measuring $ZI$ and $IZ$ and classically adding their measurement outcomes. Another unique property of quantum systems is that when two systems are placed in certain joint states, even separable ones, one cannot find a local measurement which leaves all correlations undisturbed. For instance, there exists no local measurement acting solely on Alice or Bob that would not alter the separable state $\rho = \frac{1}{4} \left( \ket{ge}\bra{ge} + \ket{eg}\bra{eg} + \ket{-+}\bra{-+} + \ket{+-}\bra{+-} \right)$, where $\ket{g}$ ($\ket{e}$) is the eigenstate of $Z$ with eigenvalue $1$ ($-1$). Any local measurement would yield outcomes both statistically uncertain due to the mixture, and quantum mechanically uncertain due to the non-existence of a measurement which is parallel to all states in this particular mixture. A result of these two properties is that if one measures $ZI + IZ$  on  $\ket{++}$, the back-action can produce non-zero components $\left<ZZ\right>$ and $\left<YY\right>$, a physically distinct result from measuring $ZI$ and $IZ$ and classically adding the outcomes. Furthermore, such a state can be quantum uncertain to every possible local observable even if it is not entangled, and thus possesses quantum correlations.
    
    The purely quantum portion of the correlations between two qubits is typically characterized by "quantum discord"~\cite{Ollivier2001,Henderson2001}, a quantity defined as the non-classical component of the mutual information. The total mutual information is defined as $I(\rho) = S(\rho^\text{A})+S(\rho^\text{B})-S(\rho)$, where $S(\rho)=- \text{Tr} \left( \rho \log_2 \rho \right)$ is the Von Neumann entropy, and where A and B refer to the "Alice" and "Bob" subsystems. The classical portion of the mutual information is defined as $J_\text{A} = S(\rho^\text{B})- \min_{ \left\{ \Pi^\text{A}_i \right\} } S( \rho^\text{B} | \left\{ \Pi^\text{A}_i \right\} )$ where $\min_{ \left\{ \Pi^\text{A}_i \right\} } S( \rho^\text{B} | \left\{ \Pi^\text{A}_i \right\} )$ is the Von Neumann entropy of B after a local measurement with projectors $\Pi^\text{A}_i$ acting on A, minimized over all possible projective measurements on A. Together, from $J_\text{A}$ and $I$, we define the discord relative to A as $\mathcal{D}_\text{A}(\rho) = I(\rho)-J_\text{A}(\rho)$. Note that we can also define a discord relative to B by exchanging the roles of A and B. 
    
    While states without discord occupy a volume of measure zero in the space of all possible mixed two-qubit states, such states are strongly favored in nature. This is due to the tendency of the environment to be coupled to local observables only. Environmental monitoring results in projection into eigenstates of local operators to which the environment is coupled, a process sometimes called Einselection~\cite{Zurek2003}. Einselection works not only to suppress entanglement, but also to suppress discord. On the other hand, applying a non-local joint measurement to a non-discordant state can produce discord. More discord of the final state implies a stronger dominance of non-local joint measurement relative to environmental monitoring of local properties or imperfections in our engineered non-local joint measurement. Discord is thus the proper quantity to characterize the effect of non-local joint measurement that produces non-classical mutual information, but not a provably entangled state. While in principle discord can be created between two remote systems by local operations suplemented by classical communications, it is of value when such classical means of communication are not available or impractical.
    
    In our experiment, we utilize the circuit-QED architecture~\cite{Blais2004,Wallraff2004} to implement a non-local joint measurement on two non-communicating systems and characterize the generated two-qubit correlations. As shown in Fig. 1a, Alice and Bob are two non-communicating qubit-cavity systems connected to the signal and idler ports of a Josephson Parametric Converter (JPC)~\cite{Bergeal2010,Bergeal2010a,Abdo2011,Roch2012}, which serves as a $ZI+IZ$ measurement apparatus while being blind to single-qubit information. This measurement is achieved by the following process: When both qubit-cavity systems are driven simultaneously with tailored pulses of coherent microwave light, the states of the qubits are entangled with these pulses, and the latter are then added and amplified by the JPC. With proper choice of pulse envelopes and drive phases on Alice and Bob (see Fig.1c), the JPC yields identical measurement outcomes for the states $\ket{eg}$ and $\ket{ge}$, indicating the presence of a single excitation in the two qubit system, but without revealing which subsystem holds this excitation. The remaining two measurement outcomes associated with pointer states $\ket{gg}$ and $\ket{ee}$ report the presence of zero or two excitations, and are discarded in discord production. This measurement process is described fully by a Stochastic Master Equation which has been solved analytically~\cite{Silveri2016}. With this Stochastic Master Equation, it is possible to calculate the optimal drive fields and filtering which we utilized in the experiment. Before examining these in detail, we describe our experiment setup.

\begin{figure}
\centering
\includegraphics[width=0.9\linewidth]{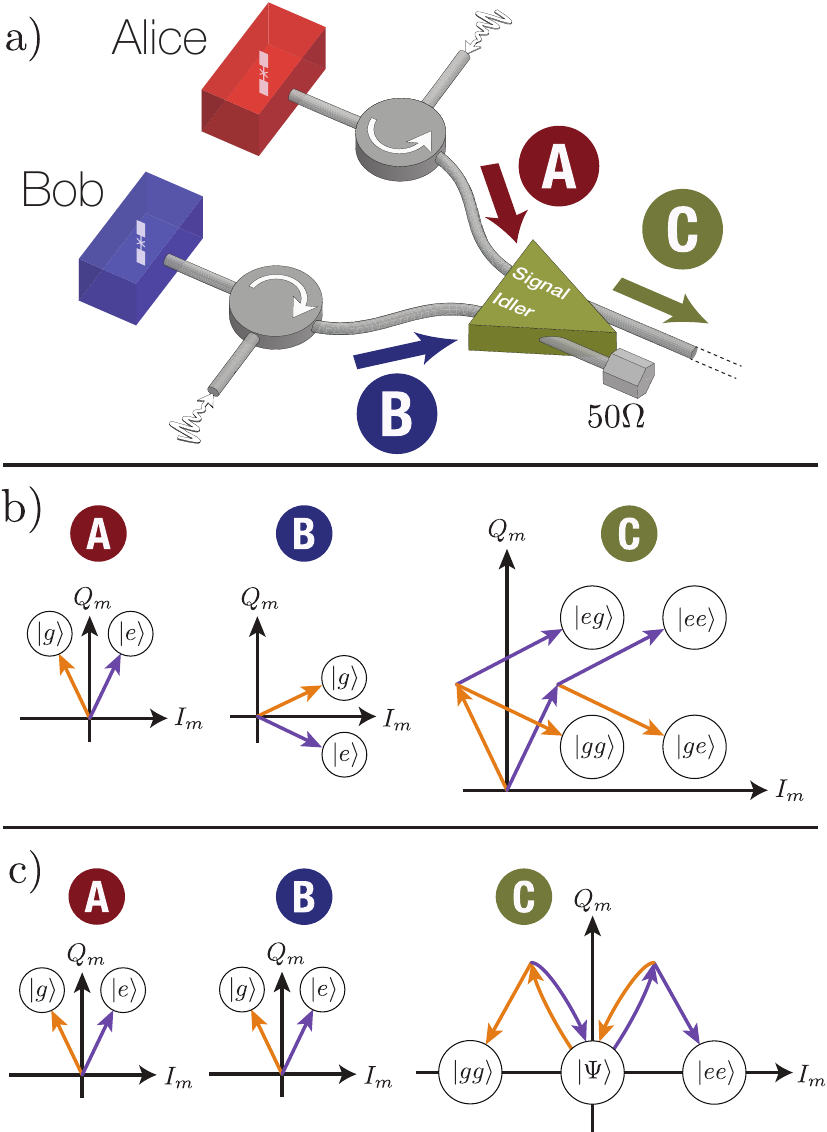}
\caption{\label{JPC_to_BS}
 (a)~Schematic representation of the experiment. Readout pulses of microwave light are simultaneously delivered to two superconducting 3D-transmon~\cite{Paik2011}
 qubits Alice and Bob through microwave circulators. After Alice and Bob impart state-dependent
 dispersive shifts to the pulses visiting them, the JPC sums and amplifies these pulses. The JPC signal and idler outputs are complex conjugates of each other and thus contain the same information. The superfluous idler output is terminated in a cold load, while the signal output is monitored. Varying the relative phase between drive pulses selects different measurement eigenstates. (b) Driving Alice and Bob $\pi/2$ out of phase results in a local joint measurement with eigenstates $\Ket{gg}$, $\Ket{ge}$, $\Ket{eg}$,
 $\Ket{ee}$. Note that the JPC signal output field contains the complex conjugate of the amplified field from Bob, as the field undergoes a complex conjugation in conversion. (c) When Alice and Bob are driven in phase, the measurement is associated with eigenstates $\Ket{gg}$, $\Ket{ee}$, and the manifold of odd Bell states
 $\Ket{\Psi_\phi}=\frac{1}{\sqrt{2}}\left(\Ket{eg}+e^{\text{i}\phi}\Ket{ge}\right)$, implementing a non-local joint measurement.
 In this odd Bell manifold, which-qubit information erasure is accomplished through the indistinguishably of the responses associated with $\left|ge\right>$ and
 $\left|eg\right>$, arising from the cancellation of phase shifts in the JPC. Note that some arrows have been presented curved for clarity.
}
\end{figure}

\section{Two Qubit Experiment Setup}

The experiment, anchored to a dilution refrigerator below 20 mK, consists of the two Alice and Bob superconducting transmon qubits ($\omega_{\text{A},\text{q}}/2\pi=4.9\; \text{GHz}$ and $\omega_\text{\text{B},\text{q}}=4.5\; \text{GHz}$), in separate 3D cavities. The cavities are each connected to the signal (A) and idler (B) ports of a JPC through two circulators as shown in Fig. \ref{JPC_to_BS}a (see Fig.~\ref{DIA} for a detailed schematic). The two-qubit measurement is performed by simultaneously driving Alice's and Bob's cavities at their respective readout frequencies ($\omega_{\text{A},\text{c}}/2\pi=7.48\; \text{GHz}$ and $\omega_\text{\text{B},\text{c}}/2\pi=9.15\; \text{GHz}$), creating flying coherent states with phases entangled with their respective qubits. The JPC sums and amplifies Alice's and Bob's cavity output fields. This output was further amplified by a cryogenic HEMT amplifier before subsequent room temperature amplification, demodulation, and digitization. Though duplicate copies of this output appear on both ports of the JPC, we elect to monitor the lower frequency signal port (C) since it falls within the bandwidth of the lowest noise temperature HEMT amplifier commercially available.

	Proper summation of readout signals requires that the frequencies of the resonators inside the JPC be tuned to match the corresponding readout cavity frequency. This condition is met to better than 200~kHz by tuning the frequency of each cavity with a screw, and tuning the JPC resonator frequencies with an applied magnetic flux. Alice and Bob in our experiment have differing cavity bandwidths $\kappa_\text{A} / 2 \pi = 5.1\; \text{MHz}$, $\kappa_\text{B} / 2 \pi = 3.8\; \text{MHz}$ and dispersive shifts $\chi_\text{A} / 2 \pi = 3.8 \; \text{MHz}$, $\chi_\text{B} / 2 \pi = 1.8\; \text{MHz}$. Differing cavity parameters implies that Alice and Bob will process their drive pulses differently. While seemingly fatal to the indistinguishability of the Alice and Bob signals, these discrepancies were compensated by shaping of temporal modes (see next section).

	We operated the JPC as a phase-preserving amplifier by supplying it with a pump drive at the sum of the internal signal and idler resonator frequencies. In the high gain limit and with proper choice of pump phase, the in-phase ($I$) quadrature and out-of-phase ($Q$) quadrature components of the output field on port C are related to the corresponding input quadratures by the relation $I_C = I_A + I_B$ and $Q_C = Q_A - Q_B$. With this relation, two kinds of joint measurements are implemented, depending on the phases of the drives applied to the two cavities.
    
    The first kind, which we term the local joint measurement, is realized by driving the cavities 90 degrees out of phase as shown in Fig.~\ref{JPC_to_BS}b. The JPC sums the two incoming fields to yield four outcome distributions in the $IQ$ plane corresponding to the four product states. Thus, in the projective limit, $I_C$ encodes the result of measuring the observable $IZ$ while $Q_C$ encodes the result of measuring the observable $ZI$. This kind of joint measurement was used in our experiment to perform tomography.
    
    On the other hand, a second kind of joint measurement, which we term the non-local joint measurement, is obtained by driving the cavities in phase. As shown in Fig.~\ref{JPC_to_BS}c, the measurement results in overlap of outcome distributions corresponding to $\ket{ge}$ and $\ket{eg}$. This yields three measurement outcomes, where the overlapping $\ket{ge}$ and $\ket{eg}$ outcome distributions no longer betray the location of the associated single excitation. This mode of operation is blind to single-qubit information. In the projective limit, $I_C$ encodes the result of the measurement observable $IZ + ZI$. For $I_C$ near the origin, the corresponding eigenstates are the manifold of odd parity Bell states.

The outcome of the non-local joint measurement along the $Q_C$ quadrature must also be recorded as it results in a non-trivial back-action. During a measurement, the cavity photon population fluctuates randomly due to photon shot noise, resulting in stochastic phase kicks to the qubit from the AC Stark effect~\cite{Schuster2005}. If continuously monitored, these phase kicks do not harm the purity of the qubits~\cite{Hatridge2013}. Crucially, in the two-qubit case, not only should we engineer a measurement which is blind to the origin of single-qubit $Z$ information, but also to the origin of the information about cavity population during the measurement. As shown in Ref.~\onlinecite{Silveri2016}, when outcome $I_C$ is recorded near the origin, the outcome $Q_C$ determines the value of the global phase $\phi$ of $\ket{ge}+e^{i\phi}\ket{eg}$ within the odd-parity manifold which would be the resulting state. Thus, the amplifier output $I$ quadrature tells us which parity the state has, while the amplifier output $Q$ quadrature tells us which phase it has.

\section{Maximal Erasure of Single-Qubit Information: Temporal Mode Matching}

With differing cavity parameters, identical drive pulses will result in differing temporal envelopes of the readout signals coming from Alice's and Bob's cavities. These differing envelopes will betray the origin of single-qubit information after summation. In the limit of long measurement time, i.e. $T_m \gg \kappa^{-1}, \chi^{-1}$, we need only to adjust the relative measurement strengths such that the $\ket{ge}$ and $\ket{eg}$ outcome distributions in the amplifier output phase space overlap in steady state. However, when $T_m$ becomes comparable to the cavity information ring-up and ring-down time, determined by $\kappa^{-1}$ and $\chi^{-1}$, the ring-up and ring-down portions of the pulse have a non-negligible weight. Therefore, the wavepacket processed by the JPC and associated with measurement outcomes $\ket{ge}$ and $\ket{eg}$ will differ significantly. However, it is possible to compensate for this effect up to a vertical shift in the $IQ$ plane, which corresponds only to a global phase shift to the final state. We engineered drive pulses to ensure that the wavepackets at the output of the JPC associated with $\ket{ge}$ and $\ket{eg}$ are identical. This type of pulse engineering is called temporal mode matching.

    During a dispersive measurement, the outgoing field emitted by the cavity becomes entangled with the qubit. We write the combined state of the outgoing field and the qubit as $\ket{\{\alpha_g(t)\}}\otimes\ket{g} + \ket{\{\alpha_e(t)\}}\otimes\ket{e}$, where $\ket{\{\alpha_i(t)\}}$ denotes the state of the outgoing field referred to its value as it was traversing the cavity at time $t$. The curly brackets inside the ket allude to the multiplicity of degrees of freedom of the transmission line in which the outgoing field propagates. These signals are then processed by the JPC resulting in filtered signals $\alpha'_{g,e}(t) = H(t) \star \alpha_{g,e}(t)$ where $H(t)$ is the response function of the JPC and $\star$ denotes convolution. Here, $H(t)$ is normalized: $\int_{-\infty}^{\infty} dt\, \left| H(t) \right|^2 = 1$.
    
    Let us now introduce the efficiencies $\eta_{\text{A}}$ and $\eta_{\text{B}}$ of the measurement chains of Alice and Bob, respectively. These parameters affect the amount of single-qubit $Z$ information obtained by the measurement. This information is determined by the difference $\mathcal{S}_i(t)=\sqrt{\eta_i \kappa_i}\left(\alpha'_{i,g}(t)-\alpha'_{i,e}(t)\right)$, with $i=\text{A},\text{B}$ indexing Alice's and Bob's channels, where the noise standard deviation per quadrature is set to be the natural single mode quantum fluctuation $\sigma_q=1/2$~\cite{Gambetta2006}. As shown in Appendix~\ref{app:tmm}, by matching $\mathcal{S}_i$ for the Alice and Bob channels for all time by suitable design of cavity drive pulses, the measurement is made insensitive to single-qubit information~\cite{Silveri2016}. We satisfy this temporal mode matching condition by choosing a JPC output wavepacket for the $\{\ket{ge},\ket{eg}\}$ manifold, inverting the response functions of the JPC and cavity, and calculating what cavity drive pulse would yield this desired JPC output wavepacket. The choice of this JPC output wavepacket is in principle arbitrary, though a narrow bandwidth JPC output wavepacket will correspond to narrow bandwidth cavity drive pulses. Such a signal will be more efficiently processed by the JPC, but is in conflict with the relatively short lifetimes of the qubits. We discuss the precise tailoring of the JPC output wavepacket in Appendix~\ref{app:tmm} and depict the resulting temporally matched waveforms in Fig.~\ref{TMM}.
    
\section{Experimental Protocol For Two Qubit Back-Action Characterization}

\begin{figure}
\centering
\includegraphics[width=0.9\linewidth]{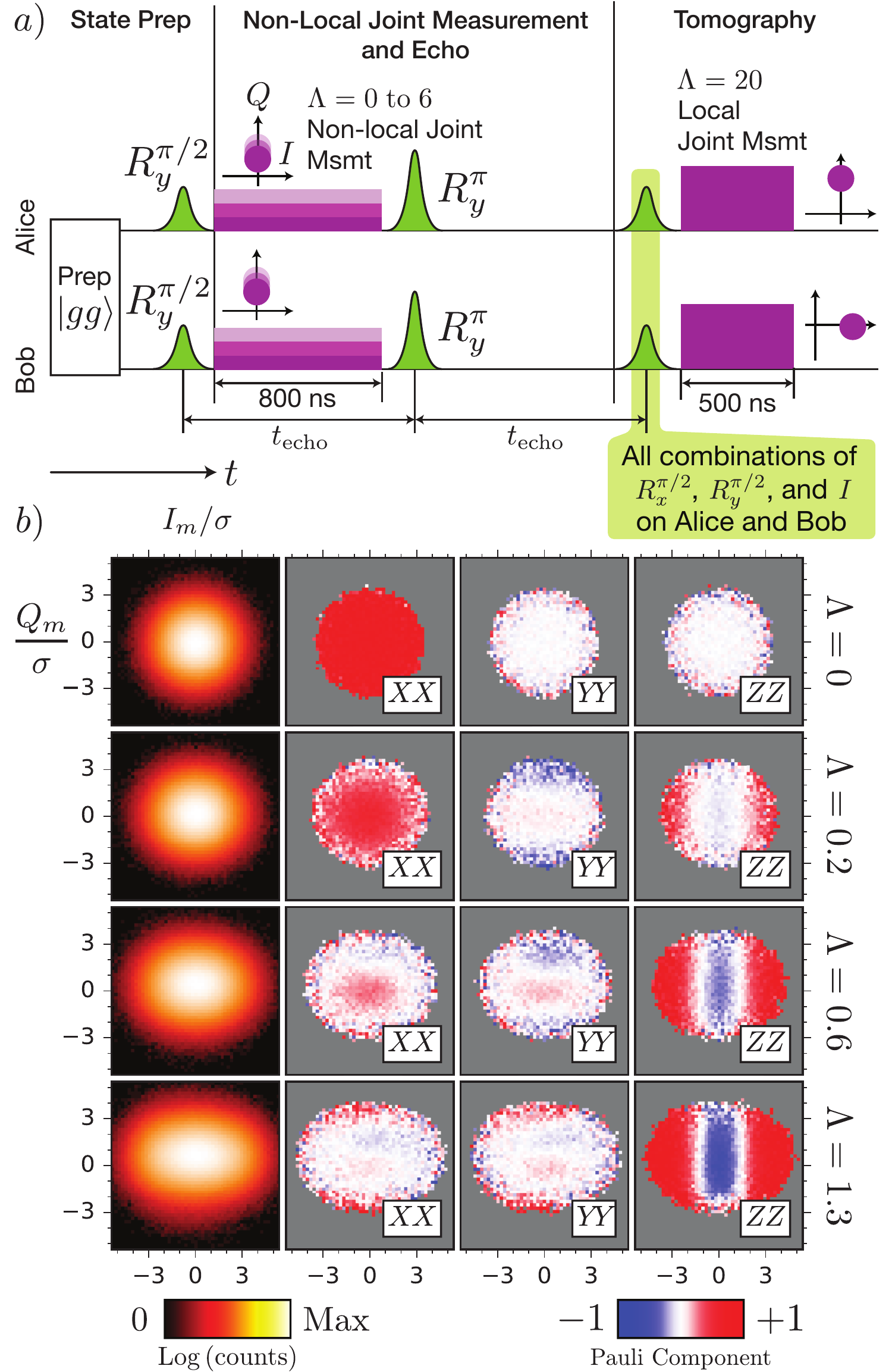}
\caption{\label{TOMO}(a)~Pulse sequence for characterizing which-qubit information erasure through measurement back-action. The qubits are initialized in $\left|gg\right>$ and then rotated to the product state $\Ket{\psi_i}=\frac{1}{2}\left( \Ket{g} + \Ket{e} \right) \otimes \left( \Ket{g} + \Ket{e} \right)$.
A variable strength non-local joint measurement is applied by driving both cavities such that the measurement outcomes associated with $\Ket{ge}$ and $\Ket{eg}$ are indistinguishable. We then apply $\pi$ pulses on both qubits and wait a time $t_{echo}$ equal to the time elapsed since the first $\pi/2$ pulse. Two-qubit tomography is then performed by applying all combinations of $R_x^{\pi/2}$, $R_y^{\pi/2}$, and $I$ and then performing a two qubit local joint measurement. The joint measurement provides two bits of information, giving each single-qubit expectation value
$\left<XI\right>$, $\left<IX\right>$, $\left<YI\right>$, ... as well as parity expectation values
$\left<ZZ\right>$, $\left<ZX\right>$, $\left<XZ\right>$, ...
(b)~Histograms of measurement outcomes $(I_m, Q_m)$ and conditional maps of Pauli components versus measurement outcome performed at measurement strengths from $\Lambda=0$~to~$1.3$.
}
\end{figure}

We now describe the experiment we have performed to characterize the two-qubit correlations generated by the mode matched non-local joint measurement. As shown in Fig. \ref{TOMO}a, the Alice and Bob qubits are first reset to the ground state, and then are rotated to the $\ket{+}$ state. Alice's and Bob's cavities are simultaneously driven by properly tailored readout pulses, with an overall amplitude scale factor, thus implementing a variable strength, mode matched non-local joint measurement. The digitized, demodulated output field of the JPC is then integrated with the optimal weighting function $\alpha'_{g}(t)-\alpha'_{e}(t)$ (where the choice of drive pulses has yielded identical $\alpha'_g-\alpha'_e$ for both cavities), resulting in the quantity $I_m + \text{i} Q_m$ which fully characterizes the outcome of the measurement. This weak non-local joint measurement is followed by one of the full set of two-qubit tomography operations, recording, at the level of the ensemble of measurements, the two-qubit state. This set consists of all combinations of $I$, $R_x^{\pi/2}$, and $R_y^{\pi/2}$ unitaries on both qubits, the sequence ending by a strong simultaneous measurement of both qubits (local joint measurement). From the outcomes of these nine pairs of one qubit measurements, we reconstruct all expectation values of the 16 Pauli operators of the two-qubit system (including $II$), all of which are tagged by the corresponding non-local measurement outcome $I_m + \text{i} Q_m$ of flying signal information. 

    In Fig. \ref{TOMO}b, we show histograms of the non-local joint measurement $I_m + \text{i} Q_m$ as a function of the measurement strength $\Lambda=\int_{-\infty}^\infty dt\, \left|\mathcal{S}_i(t)\right|^2$ (first column). In the remaining columns of the figure, the dependence of the average of three selected two-qubit Pauli operators $XX$, $YY$, and $ZZ$ are shown as a function of their corresponding $(I_m,Q_m)$ non-local measurement outcome. In this part of the figure, the $IQ$ plane has been tiled into $51 \times 51 = 2601$ square bins, and the corresponding two-qubit Pauli component have been calculated by averaging only those tomography results whose non-local measurement outcome $(I_m, Q_m)$ was contained within the corresponding bin. The protocol was repeated for each measurement strength for a total of $4.5 \times 10^6$ shots. We refer to these panels as conditional tomograms.
    
    The first row features data taken at the measurement strength $\Lambda=0$, corresponding to no non-local measurement taking place. The two qubits are expected to remain in the state $\ket{++}$ (the state the Alice and Bob qubits have been prepared in), implying $\left<XX\right>=1$, $\left<YY\right>=0$, $\left<ZZ\right>=0$. With increasing measurement strength, the measurement process projects the qubits into a state of definite parity along $Z$, that is $\left<ZZ\right> \neq 0$. This behavior is shown by the $\left<ZZ\right>$ conditional tomogram as a function of measurement strength. With increasing measurement strength, outcomes near  $I_m\sim0$ project the qubits into the $\{\ket{ge},\ket{eg}\}$ manifold, corresponding to $\left<ZZ\right> = -1$. Similarly, large positive (negative) $I_m$ outcomes project the qubits into $\ket{gg}$ ($\ket{ee}$) respectively, corresponding to $\left<ZZ\right> = +1$. A complementary behavior is observed in the XX and YY conditional tomograms as a function of measurement strength. These show sinusoidal fringes as a function of the $Q_m$ outcome. The value of $Q_m$ is proportional to the relative phase between $\ket{ge}$ and $\ket{eg}$ for the the final two-qubit state. The amplitude of these fringes is expected to be sensitive to the inefficiency of the measurement process, which dephases superpositions between $\ket{ge}$ and $\ket{eg}$. This effect is shown by the reduction in fringe contrast with increasing measurement strength.
    
\begin{figure*}
\centering
\includegraphics[width=0.9\linewidth]{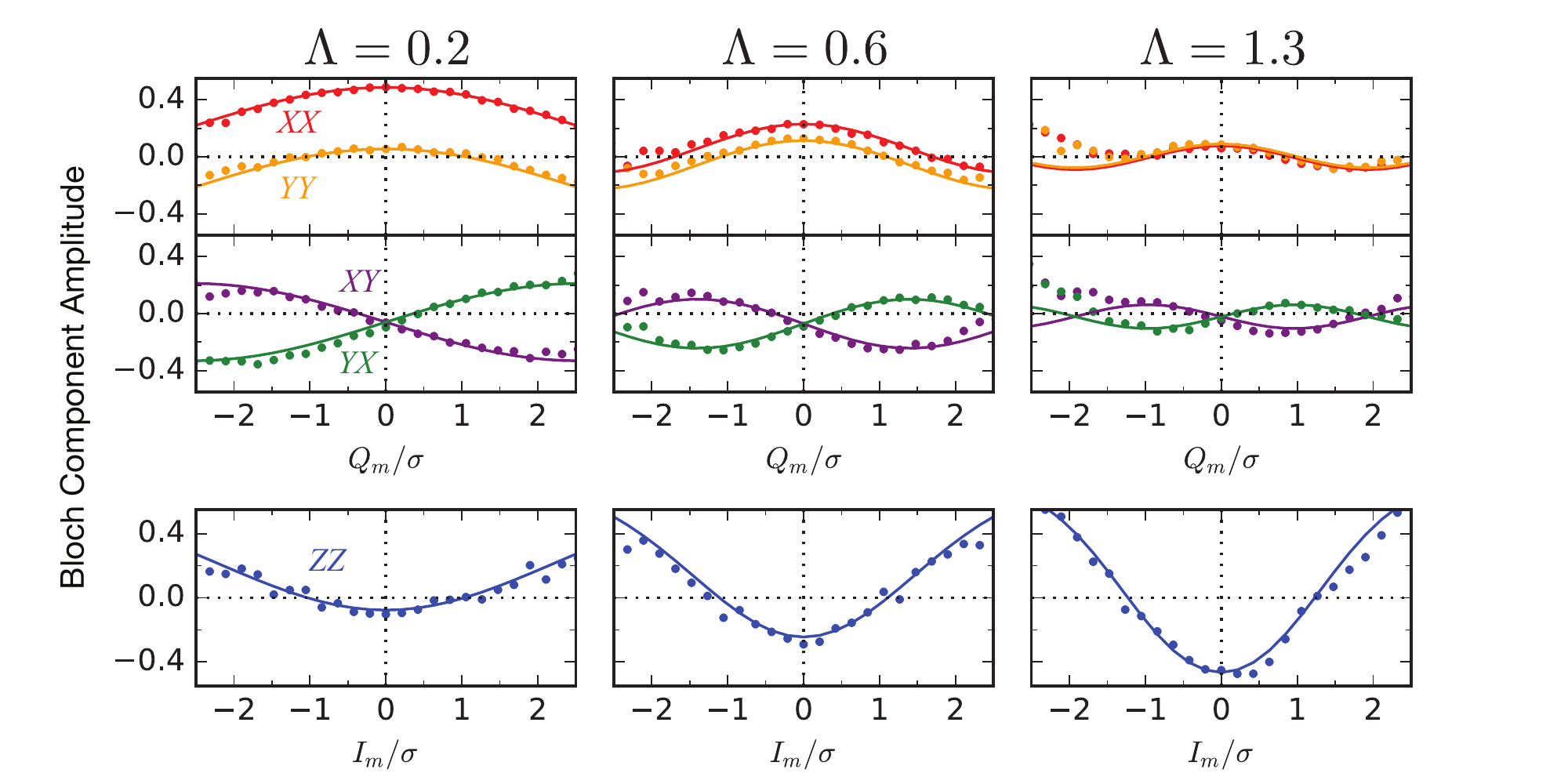}
\caption{\label{SLICES}
Cuts of Fig.~\ref{TOMO}b at $I_m=0$ (upper panel) and $Q_m=0$ (lower panel) along with theory plots. Increasing measurement strength purifies out $\Ket{gg}$ and $\Ket{ee}$ as the outcome distributions associated with these states are forced further out along $I_m$, demonstrated by increasing negativity of the $\left<ZZ\right>$ component at $I_m=0$. Increasing measurement strength also results in reduced purity in the odd Bell-manifold due to measurement inefficiency. These cuts display good agreement with the theory in~\cite{Silveri2016}. Information on how the parameters were determined is available in Appendix~\ref{app:model}. Sudden rise in Pauli components in top two plots near negative $Q_m/\sigma$ is due to $\ket{f}$ state poisoning.
}
\end{figure*}

    Taking cuts along $I_m=0$ for $\left<XX\right>$, $\left<XY\right>$, $\left<YX\right>$, and $\left<YY\right>$ and along $Q_m=0$ for $\left<ZZ\right>$, and comparing them to theory from Ref.~\onlinecite{Silveri2016} leads to the data shown in Fig.~\ref{SLICES}. The theory curves incorporate as parameters the intrinsic dephasing ($T_2$ of both qubits), the accumulated dephasing and AC-Stark shift due to the non-local measurement, the measurement efficiencies $\eta_{\text{A,B}}$, as well as the measurement strength (see Appendix~\ref{app:model}). All these parameters are determined through separate characterization experiments. The excellent agreement between data and theory demonstrates that the theoretical description given in Ref.~\cite{Silveri2016} captures the details of our experiment.

\section{Quantum discord analysis}
\begin{figure}
\centering
\includegraphics[width=\linewidth]{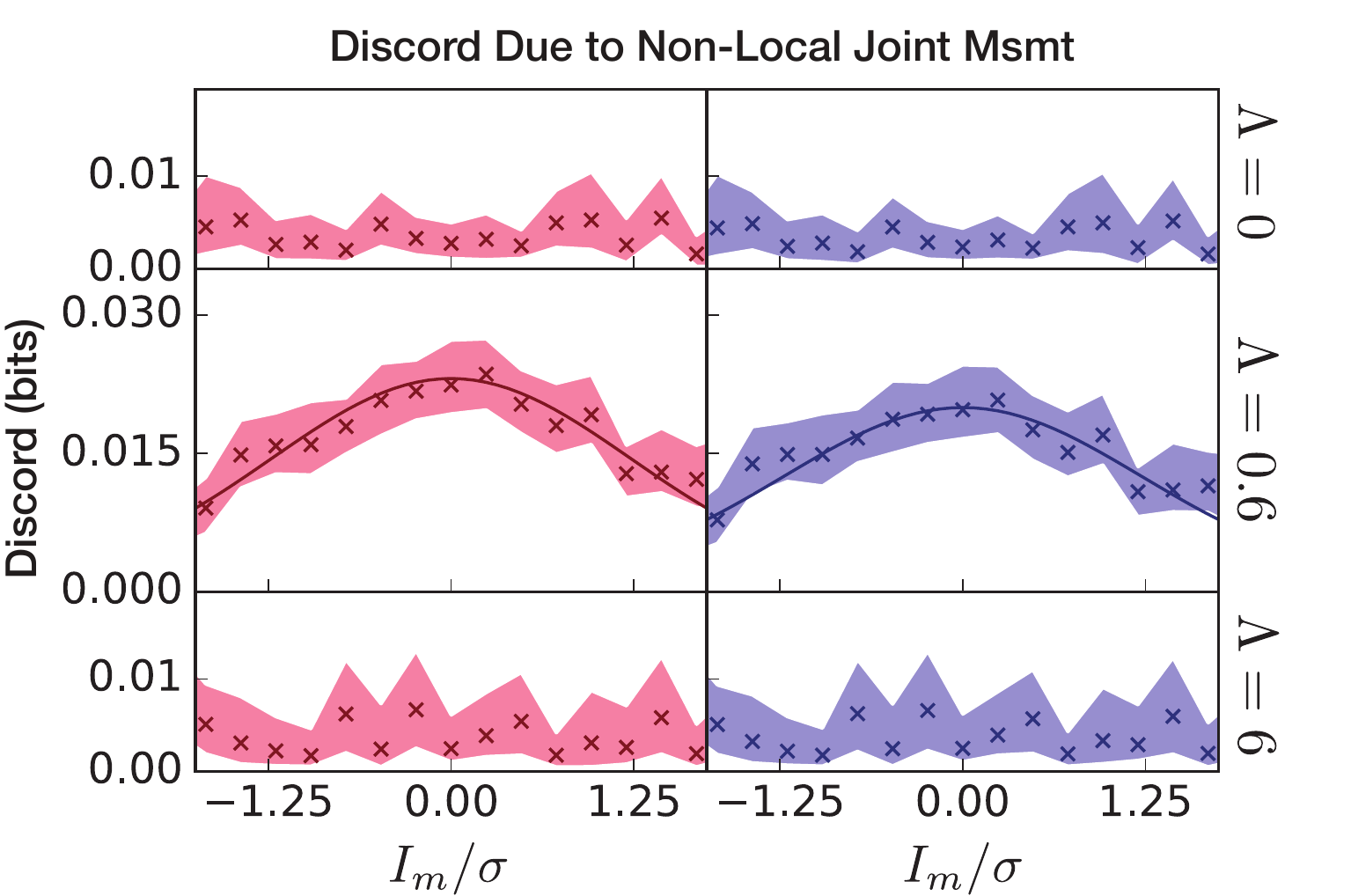}
\caption{\label{DISCORD}
Discord of system after application of a variable strength non-local measurement ($\times$) and 95\% confidence interval (band) determined through bootstrap (see Appendix~\ref{app:discordcalc}), displayed with respect to Alice (red) and Bob (blue). Discord is negligible at measurement strength $\Lambda = 0$ due to lack of discord in the initial state $\ket{++}$. Discord is also negligible at high measurement strength $\Lambda = 6$ since the corresponding odd mixed state also lacks discord. The curve for $\Lambda = 0.6$ is numerically calculated from the expected density matrix for this strength given by the analysis in Ref.~\onlinecite{Silveri2016}
}
\end{figure}

We now proceed to analyze the amount of quantum discord generated by the non-local measurement (see Appendix~\ref{app:discordcalc}) for full details of the procedure). Calculation of quantum discord requires a positive semi-definite density matrix of trace one, which we estimate using a maximum likelihood reconstruction procedure~\cite{Banaszek1999}.  This procedure uses all tomography results associated with a given $(I_m, Q_m)$ bin to construct the most likely density matrix associated with that particular bin. Next, since as predicted by theory, all states in a single $I_m$ column are related up to local rotations, we average the reconstructed density matrices along the $Q_m$ axis after applying such a rotation. There is however the caveat that the discord of the convex combination of two states may have higher discord than either of the states individually, and thus this averaging procedure could counter-intuitively increase discord. However, we do not expect our procedure to increase discord in such a way, due to reasons described in Appendix~\ref{DISCORD}.

The results of the calculated discord are shown in Fig.~\ref{DISCORD}. We calculated both discord relative to Alice and relative to Bob. The relatively small difference between discord relative to Alice and discord relative to Bob is attributed to differing measurement efficiencies of channels associated with Alice's and Bob's cavities. Due to the small amount of discord generated by the experiment, we applied a bootstrapping procedure to estimate the error due to sampling and display the corresponding error bands~\cite{Efron1993}. This procedure reconstructs 2000 bootstrapped density matrices per $\left(I_m,Q_m\right)$ pixel by sampling tomography results with replacement.  As expected, the discord is negligible at weak measurement strengths since the qubits remain in the product state $\ket{++}$.  Similarly at large measurement strengths, the qubits will be dephased into a fully mixed state $\frac{1}{2}\left( \ket{ge}\bra{ge} + \ket{eg}\bra{eg} \right)$, which also has no discord. The lack of discord at these extremes is strong evidence that systematic errors have not inflated the amount of discord present in the system. On the other hand, for moderate measurement strengths, the non-local measurement results in qubit states with finite discord. Discord peaks when the measurement outcomes are near $I_m\sim0$ corresponding to states in the $\{\ket{ge},\ket{eg}\}$ manifold. Discord falls off away from $I_m\sim0$, as these measurement outcomes herald the even product states, which also have no discord. While no analytical form for discord is known, we can use the theoretical predictions for the final qubit density matrix that we have verified in the previous section, to numerically calculate discord for every $I_m$ value. This numerically calculated curve, also shown in Fig.~\ref{DISCORD} agrees well with the values extracted from the experiment. 

\section{Outlook and Conclusions}

We have successfully generated discord by measuring a joint non-local property of two remote systems. This demonstrates that this joint measurement is truly blind to local properties of these systems. While discordant states could in principle be produced through local operations and classical communication, there is no deliberate communication process between our two qubits taking place, and thus our experiment illustrates a novel protocol in which discord is generated. We believe that with improvement of experimental hardware, our experiment could be used as one of the simplest possible schemes for concurrent continuous variable remote entanglement in circuit-QED. Due to its ability of bridging the resonant frequency difference between systems, this protocol features excellent isolation between subsystems, making this scheme appealing for a modular quantum architecture~\cite{Monroe2014a}. The experimental setup is sufficiently well characterized to demonstrate that the quantum correlations are limited by the inefficiency of the components in the first stage of the measurement. This efficiency improvement might be achieved through use of an amplifier that requires a single-ended rather than a differential coupling which would remove the need for the lossy microwave hybrids present in the JPC~\cite{Liu2017}. Further improvements in efficiency might also be achieved through the use of directional amplification~\cite{Sliwa2015,Lecocq2017}, which would remove the need for lossy microwave circulators.

Beyond the efficiency threshold for entanglement in this experiment lies the prospect of an extremely simple loophole-free Bell test~\cite{Hensen2015,Shalm2015,Giustina2015}. With a quantum efficiency of $\eta_{\text{A},\text{B}}=97\%$, it is possible to obtain a Bell test violating state by non-local joint measurement. Since the phase of this Bell state is chosen by vacuum fluctuations, it is provably random. While the non-local measurement generates a continuum of Bell state phases, it would be possible to post-select those applicable to a CHSH test and demonstrate loophole-free Bell violation, as was done in Ref.~\cite{Hensen2015}.

\begin{acknowledgements}
The authors thank Mazyar Mirrahimi and Davide Girolami for helpful discussions
and Shantanu Mundhada for assistance with the experiment. We thank the Yale Center for Research Computing for guidance and use of the research computing infrastructure. Facilities use was supported by the Yale Institute for Nanoscience and Quantum Engineering (YINQE), the National Science Foundation (NSF) MRSEC DMR-1119826, and the Yale
School of Engineering and Applied Sciences cleanroom. This research was supported by the U.S. Army Research Office (Grant No. W911NF-14-1-0011), and the Multidisciplinary University Research Initiative through the U.S. Air Force Office of Scientific Research (Grant No. FP057123-C).
\end{acknowledgements}

\appendix

\section{Temporal Mode Matching and Choice of Drive Pulse}\label{app:tmm}

We first define the classical trajectory of the Alice cavity (Bob
cavity) when the Alice qubit (Bob qubit) is conditioned in $\left|g\right\rangle $
and $\left|e\right\rangle $ as $\alpha_{g}^{\text{A}}$$\left(t\right)$
and $\alpha_{e}^{\text{A}}\left(t\right)$ ($\alpha_{g}^{\text{B}}\left(t\right)$
and $\alpha_{e}^{\text{B}}\left(t\right)$). We additionally define
$H_{\text{A}}\left(t\right)$ ($H_{\text{B}}\left(t\right)$) to be
the JPC transfer function from its signal input port, connected to the Alice cavity (idler input port, connected to the Bob cavity)
to the output of the JPC. The flying field leaving the JPC, conditioned on the
qubit state for Alice $i=\ket{g},\ket{e}$ and for Bob $j=\ket{g},\ket{e}$, will then be

\begin{equation*}
\begin{split}
\Sigma_{ij}\left(t\right) &=\sqrt{\kappa_{\text{A}}\eta_{\text{A}}}\left[H_{\text{A}}\left(t\right)\star\alpha_{i}^{\text{A}}\left(t\right)\right] \\
&+\sqrt{\kappa_{\text{B}}\eta_{\text{B}}}\left[H_{\text{B}}\left(t\right)\star\alpha_{j}^{B}\left(t\right)\right]^{*}.
\end{split}
\end{equation*}

A measurement which erases which path information for the odd parity
states requires that there be no way to distinguish between measurement
outcomes $\left|ge\right\rangle $ and $\left|eg\right\rangle $,
and thus we have the condition $\Sigma_{ge}\left(t\right)=\Sigma_{eg}\left(t\right)$,
or

\begin{equation}
\begin{split}\label{matchingcondn}
\sqrt{\kappa_{\text{A}}\eta_{\text{A}}}\left[H_{\text{A}}\left(t\right)\star\alpha_{g}^{\text{A}}\left(t\right)\right]
+\sqrt{\kappa_{\text{B}}\eta_{\text{B}}}\left[H_{\text{B}}\left(t\right)\star\alpha_{e}^{B}\left(t\right)\right]^{*} \\
=\sqrt{\kappa_{\text{A}}\eta_{\text{A}}}\left[H_{\text{A}}\left(t\right)\star\alpha_{e}^{\text{A}}\left(t\right)\right]
+\sqrt{\kappa_{\text{B}}\eta_{\text{B}}}\left[H_{\text{B}}\left(t\right)\star\alpha_{g}^{B}\left(t\right)\right]^{*}.
\end{split}
\end{equation}

The complex trajectories $\alpha_{g}^{\text{A}},\alpha_{e}^{\text{A}},\alpha_{g}^{\text{B}},\alpha_{e}^{\text{B}}$
are determined by the drives applied to the
respective cavities as well as their couplings and dispersive shifts. In the Fourier domain, we have

\begin{equation}
\begin{split}\label{classicalpaths}
\alpha_{g}^{\text{A}}\left[\omega\right] &= \frac{\kappa_{\text{A}}\epsilon_{\text{A}}\left[\omega\right]/2}{-\frac{\kappa_{\text{A}}}{2}+i\left(-\frac{\chi_{\text{A}}}{2}-\omega\right)}, \\
\alpha_{e}^{\text{A}}\left[\omega\right] &= \frac{\kappa_{\text{A}}\epsilon_{\text{A}}\left[\omega\right]/2}{-\frac{\kappa_{\text{A}}}{2}+i\left(\frac{\chi_{\text{A}}}{2}-\omega\right)}, \\
\alpha_{g}^{\text{B}}\left[\omega\right] &= \frac{\kappa_{\text{B}}\epsilon_{\text{B}}\left[\omega\right]/2}{-\frac{\kappa_{\text{B}}}{2}+i\left(-\frac{\chi_{\text{B}}}{2}-\omega\right)}, \\
\alpha_{e}^{\text{B}}\left[\omega\right] &= \frac{\kappa_{\text{B}}\epsilon_{\text{B}}\left[\omega\right]/2}{-\frac{\kappa_{\text{B}}}{2}+i\left(\frac{\chi_{\text{B}}}{2}-\omega\right)},
\end{split}
\end{equation}

where $\epsilon_{\text{A}}\left[\omega\right]$ ($\epsilon_{\text{B}}\left[\omega\right]$)
is the Fourier transform of the drive applied to the Alice (Bob) cavity.
Here we suppose we are driving the cavities at a frequency half way between the peaks
associated with the qubit $\ket{g}$ and $\ket{e}$ states. Thus temporal mode
matching consists of choosing $\epsilon_{A}$ and $\epsilon_{B}$
such that equation \ref{matchingcondn} is satisfied for all time, or in other words
$f_{\text{A}}\left(t\right)=f_{\text{B}}\left(t\right)$ where the Fourier transforms of $f_A$ and $f_B$ are given by

\begin{equation}\label{ffunction}
\begin{split}
f_{\text{A}}\left[\omega\right] &= \sqrt{\kappa_{\text{A}}\eta_{\text{A}}}\left[H_{\text{A}}\left[\omega\right]\left(\alpha_{g}^{\text{A}}\left[\omega\right]-\alpha_{e}^{\text{A}}\left[\omega\right]\right)\right], \\
f_{\text{B}}\left[\omega\right] &= \sqrt{\kappa_{\text{B}}\eta_{\text{B}}}\left[H_{\text{B}}\left[\omega\right]\left(\alpha_{g}^{\text{B}}\left[\omega\right]-\alpha_{e}^{\text{B}}\left[\omega\right]\right)\right].
\end{split}
\end{equation}

We furthermore model the response function of the JPC in the $s$-domain
with the one-pole response functions

\begin{equation}\label{jpcrespfn}
\begin{split}
H_{\text{A}}\left[\omega\right]=G\frac{\frac{\kappa_{\text{JPC}}}{2}+i\Delta_{\text{A}}}{\frac{\kappa_{\text{JPC}}}{2}+i\left(\Delta_{\text{A}}+\omega\right)}, \\
H_{\text{B}}\left[\omega\right]=G\frac{\frac{\kappa_{\text{JPC}}}{2}+i\Delta_{\text{B}}}{\frac{\kappa_{\text{JPC}}}{2}+i\left(\Delta_{\text{B}}+\omega\right)},
\end{split}
\end{equation}

where $\kappa_{\text{JPC}}$ is the JPC bandwidth in both reflection
and trans-gain, and $\Delta_\text{A}$ and $\Delta_\text{B}$ are the detunings between the associated cavity drive
frequency and the frequency at which the JPC gain is maximal on the
associated input port. Note that for Alice, this is the detuning between
the Alice cavity drive frequency and the frequency at which the \textbf{reflection}
gain is maximized. For Bob, this is the detuning between the Bob cavity
drive frequency and the frequency at which the \textbf{trans}-gain
is maximal, \textbf{before} conversion. Combining equations \ref{classicalpaths} to \ref{jpcrespfn} we find that $f_\text{A}\left[\omega\right]$ is given by

\begin{equation}
\begin{split}
f_{\text{A}}\left[\omega\right]  = & \sqrt{\kappa_{\text{A}}\eta_{\text{A}}}\left[H_{\text{A}}\left[\omega\right]\left(\alpha_{g}^{\text{A}}\left[\omega\right]-\alpha_{e}^{\text{A}}\left[\omega\right]\right)\right]\\
  = & G\sqrt{\kappa_{i}\eta_{i}}\left(\frac{\frac{\kappa_{\text{JPC}}}{2}+i\Delta_{\text{A}}}{\frac{\kappa_{\text{JPC}}}{2}+i\left(\Delta_{\text{A}}+\omega\right)}\right) \\
& \times \left(\frac{2i\kappa_{\text{A}}\chi_{\text{A}}}{\left(\kappa_{\text{A}}+2i\omega\right)^{2}+\chi_{\text{A}}^{2}}\right)\epsilon_{\text{A}}\left[\omega\right].
\end{split}
\end{equation}

Bob's JPC signal $f_{\text{B}}\left[\omega\right]$ directly follows from the same type of equation.
By choosing a practical narrow bandwidth function $f\left[\omega\right]$, we can invert these relations
and find, after an inverse Fourier transform, $\epsilon_{\text{A}}\left(t\right)$ and $\epsilon_{\text{B}}\left(t\right)$.

\begin{figure*}
\centering
\includegraphics[width=\textwidth]{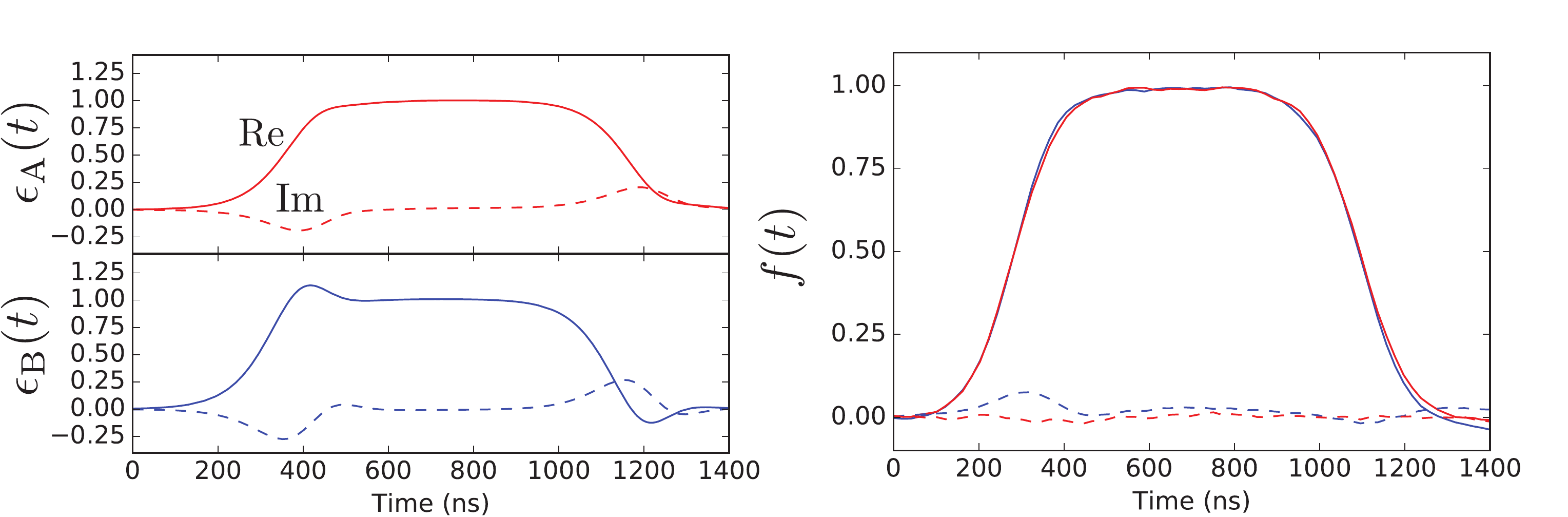}
\caption{\label{TMM}Synthesized temporal mode matching pulses (left panels) and resulting temporal envelopes (right panels). Solid lines depict the real part of the complex signal and dotted lines depict the imaginary part. Red lines are assocated with the Alice subsystem and blue lines are associated with the Bob subsystem. Left panels depict the synthesized temporal mode matching pulses, normalized to their maximum amplitudes. The right panel depicts the measured temporal envelopes which result from the use of the synthesized drive pulses.}
\end{figure*}

In the experiment, we chose $f\left[\omega\right]$ such that its Fourier transform $f(t)$ was

\begin{equation}
\begin{split}
f_{\text{A},\text{B}}\left(t\right)= & \frac{C}{2}\tanh\left(\frac{t+t_{\text{duration}}/2}{t_{\text{slew}}}\right) \\ 
&-\frac{C}{2}\tanh\left(\frac{t-t_{\text{duration}}/2}{t_{\text{slew}}}\right),
\end{split}
\end{equation}

with $t_{\text{slew}}=80\; \text{ns}$ and $t_\text{duration} = 800\; \text{ns}$. Experimental matched temporal envelopes are shown in the right panel of Fig.~\ref{TMM}, with their associated drive pulses depicted in the left panel.

Drive envelopes were synthesized using measured dispersive shifts and coupling rates
of both Alice and Bob. Amplifier bandwidth $\kappa_{\text{JPC}}$
and shift from input frequency $\Delta_{\text{A,B}}$ were measured
using a VNA. This resulted in a pulse taking up a total of
$\text{1.4}\ \mu\text{s}$ in our sequence accounting for ring-up
and ring-down. This pulse is estimated to contribute 4\% to the measured quantum efficiency, as the bandwidth of the signal leaving each cavity (full width at half max) is calculated to be 1 MHz, well within the JPC's bandwidth. We find a $2.3\%$ mean-squared error in the temporal mode matching as defined by

\begin{equation}
\sqrt{
    \frac{
        \int_{-\infty}^{\infty}dt\,|\mathcal{S}_\text{A}(t)-\mathcal{S}_\text{B}(t)|^2
    }{
        \int_{-\infty}^{\infty}dt\,|\mathcal{S}_\text{A}(t)+\mathcal{S}_\text{B}(t)|^2
    }
}.
\end{equation}

\section{Conditional Density Matrix Parameters and Model}\label{app:model}

We compare our experiment in Fig.~\ref{SLICES} to the model from~\cite{Silveri2016}.

\begin{align}
C  = & C_{T_{2,\text{Alice}}}C_{T_{2,\text{Bob}}}C_{\text{Tomo}} \nonumber \\
  & \times \exp\left[-\left(\frac{1-\eta_{\text{A}}}{\eta_{\text{A}}}+\frac{1-\eta_{\text{B}}}{\eta_{\text{B}}}\right)\frac{\Lambda}{2}\right], \\
\Lambda  = & \frac{\bar{I}_{m}^{2}}{2\sigma_{m}^{2}},\\
\left\langle XX\right\rangle = & C\frac{-e^{-\Lambda}\cos\Theta_{+}+\cos\left[Q_{m}-\Theta_{-}\right]}{e^{-\Lambda}\cosh\left(I_{m}\right)+1},\\
\left\langle YY\right\rangle = & C\frac{e^{-\Lambda}\cos\Theta_{+}+\cos\left[Q_{m}-\Theta_{-}\right]}{e^{-\Lambda}\cosh\left(I_{m}\right)+1},\\
\left\langle XY\right\rangle = & C\frac{e^{-\Lambda}\sin\Theta_{+}-\sin\left[Q_{m}-\Theta_{-}\right]}{e^{-\Lambda}\cosh\left(I_{m}\right)+1},\\
\left\langle YX\right\rangle = & C\frac{e^{-\Lambda}\sin\Theta_{+}+\sin\left[Q_{m}-\Theta_{-}\right]}{e^{-\Lambda}\cosh\left(I_{m}\right)+1},\\
\left\langle ZZ\right\rangle = & C_{\text{Tomo}}\frac{e^{-\Lambda}\cosh\left[I_{m}\right]-1}{e^{-\Lambda}\cosh\left[I_{m}\right]+1},\\
\Theta_{-} = & \bar{Q}_{m}+\xi_{\text{A}}\Lambda-\xi_{\text{B}}\Lambda,\\
\Theta_{+} = & \xi_{\text{A}}\Lambda+\xi_{\text{B}}\Lambda.
\end{align}

We have the parameters $C_{T_{2},\text{Alice}}$, $C_{T_{2},\text{Bob}}$
$C_{\text{Tomo}}$, $\eta_{A}$, $\eta_{B}$, $\xi_{\text{A}}$, and
$\xi_{\text{B}}$, which were all measured through six independent
experiments, as listed below.
\begin{itemize}
\item $C_{T_{2},\text{Alice}}$ and $C_{T_{2},\text{Bob}}$ model the single
qubit infidelity due to decoherence and state preparation errors.
We extract these from the $\Lambda=0$ case of Fig.~\ref{TOMO} in the main paper, unconditioned
by the weak measurement.
\item $C_{\text{Tomo}}$ models the errors arising from imperfect measurements
during tomography, and is extracted by performing tomography on all
four product states.
\item $\eta_{\text{A}}$ and $\eta_{\text{B}}$ model the quantum efficiency
of the total path from the respective cavities to the room temperature
measurement systems. These are extracted by the back-action characterization
protocol described in~\cite{Hatridge2013}.
\item $\xi_{\text{A}}$ and $\xi_{\text{B}}$ model the AC stark shift accrued
over our weak measurement for the measurement strength $\Lambda=1$.
These are extracted from measurements of the Ramsey fringe shift in
the presence of a weak measurement pulse.~\cite{Campagne2015}
\end{itemize}
We note that due to mismatched bandwidths and dispersive shifts, we
do not expect our outcome distributions to have zero mean in the $Q_{m}$
axis, as assumed in~\cite{Silveri2016}. This results in a measurement
strength dependent phase shift of the fringes in the $IQ$ plane,
in addition to the AC-Stark shift, which we account for by adding
the term $\bar{Q}_{m}$ for the expression for $\Theta_{-}$. The
theory curves in Fig.~\ref{TOMO} and Fig.~\ref{SLICES} in the main paper are calculated with the values given in
table \ref{params}.

\begin{table}
\centering
\begin{tabular}{*7l}    \toprule
$C_{T_{2},\text{Alice}}$ & $C_{T_{2},\text{Bob}}$ &
$C_{\text{Tomo}}$ & $\eta_{A}$ & $\eta_{B}$ & $\xi_{\text{A}}$ &
$\xi_{\text{B}}$  \\ 
0.86 & 0.85 & 0.90 & 0.53 & 0.42 & 0.27 & 1.02\\\bottomrule
 \hline
\end{tabular}
\caption{Parameters used for theory plots in Fig.~\ref{SLICES} and Fig.~\ref{DISCORD} in the main paper}\label{params}
\end{table}

\section{Calculation of Discord}\label{app:discordcalc}

\begin{figure*}
\centering
\includegraphics[width=\textwidth]{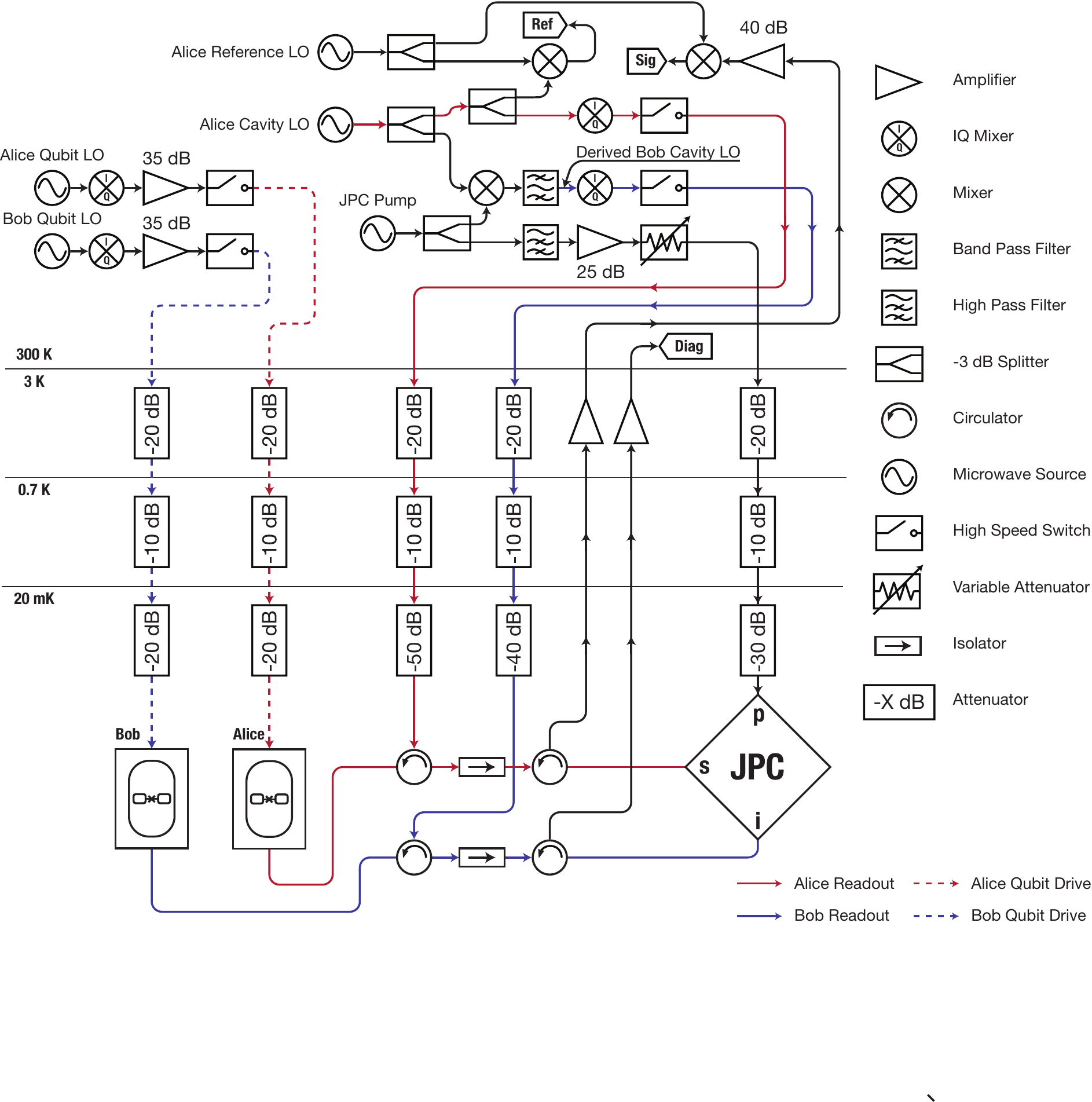}
\caption{\label{DIA} System block diagram. Modulation inputs to IQ mixers omitted for clarity. Each IQ mixer is associated with two 500 MSPS DACs, which are each attached to the I and Q inputs of the mixer for the purposes of pulse shaping. The Sig and Ref lines are monitored by 1 GSPS ADCs for capturing measurement outcomes.}
\end{figure*}

We present here a procedure for estimating the quantum discord present in our experiment. We begin by applying maximum likelihood density matrix
reconstruction~\cite{Banaszek1999} to obtain the density matrix most
likely to describe the experimental data. Crucially, this procedure
must be applied for every $I_{m}$ and $Q_{m}$ bin, as every measurement
result corresponds to a unique density matrix $\rho(I_m,Q_m)$. Additionally, since the expected
Pauli components given a certain $I_{m}$ and $Q_{m}$ also depend
on $\Lambda$, this procedure must also be repeated for every measurement
strength.

We predict that the discord associated with all values of $Q_{m}$
are identical, since all density matrices associated with identical
$I_{m}$ should be related up to local rotations. We could calculate
the discord associated with each $I_{m}$ and $Q_{m}$ bin, but due
to the nonlinearity and positivity of the discord measure, any noise
is likely to misrepresent our discord as higher than it is. We would
therefore like to average together all density matrices for a certain
$I_{m}$ after applying local rotations which maximize the purity
of an average over $Q_{m}$ values. One strategy is using the arguments
of the $Q_{m}$ dependent oscillating terms in the theory to figure
out the rotation based on a-priori experiments. This should work in
principle, but any systematic error in determining the frequency of
these fringes relative to $Q_{m}$ will reduce the purity of the marginalized
density matrix. We instead apply a numerical optimization to try and
find the correct single qubit rotations. We define a real vector $\Xi=\left(\Xi_{\text{A}},\Xi_{\text{B}}\right)$,
which links the $Q_{m}$ measurement outcome to the stochastic phase shift on Alice and Bob,
and a corresponding rotated density matrix

\begin{equation}
\rho_{\Xi}(I_m,Q_m)=U_{\Xi}\left(Q_{m}\right)\rho\left(I_{m},Q_{m}\right)U_{\Xi}^{\dagger}\left(Q_{m}\right),
\end{equation}

where the unitary rotation operator

\begin{equation}
U_{\Xi}\left(Q_{m}\right)=e^{i\,ZI\,\Xi_{\text{A}}Q_{m}}\otimes e^{i\,IZ\,\Xi_{\text{B}}Q_{m}}.
\end{equation}

We then calculate the marginalized density matrix averaged over all $Q_{m}$ as

\begin{equation}
\rho_{\Xi}^{M}\left(I_{m}\right)=\int dQ_{m}\,P\left(Q_{m}\right) \rho_{\Xi}(I_m,Q_m).
\end{equation}

These marginalized density matrices are used to calculate the average purity for this measurement strength

\begin{equation}\label{AVG-AFTER-ROT}
\gamma_{\Xi}=\int dI_{m}\,P\left(I_{m}\right)\text{Tr}\left[\left(\rho_{\Xi}^{M}\left(I_{m}\right)\right)^{2}\right]
\end{equation}

This gives a single optimal set of parameters $\Xi_{\text{Opt}}=\text{argmax}_{\Xi}\gamma_{\Xi}$
for this measurement strength. We note that the $\Xi$ which maximizes the purity $\gamma_{\Xi}$ is equivalent to choosing the $\Xi$ which maximizes the $P(Q_m)$ weighted sum of the overlaps, $\text{Tr} [ \rho_{\Xi}(I_m,Q_m) \rho_{\Xi}(I_m,Q_m') ]$, between all states in the average, corresponding to Eq.~\ref{AVG-AFTER-ROT}. We finally use this $\Xi_{\text{Opt}}$ to calculate the maximally pure marginalized density matrix $\rho^{M}_{\Xi_{\text{Opt}}}\left(I_{m}\right)$. While the frequency of these fringes can be calculated from the measurement strength $\Lambda$, this optimization procedure ensures any systematic error in measuring $\Lambda$ does not poison the fidelity of the marginalized state. We find that the $\Xi$s we extract from the marginalization procedure match our model to within $5\%$.

This marginalization procedure could be considered a form of classical communication, and could thus increase the calculated value of discord. However, as the procedure maximizes the overlap of the states over which we average, we do not expect $\rho^{M}_{\Xi_{\text{Opt}}}\left(I_{m}\right)$ to over-represent the amount of discord present in the experiment. Indeed we find that the discord produced by our marginalization procedure is generally lower than that produced by only averaging, over $Q_m$, the discord of the states characterized by the same $I_m$ bin coordinate.

To understand the degree to which this marginalization procedure reduces the average purity of the ensemble, we compare $\gamma_{\Xi_{\text{Opt}}}$ to the average purity defined as

\begin{equation}
\gamma_{\text{Avg}}=\int dI_{m}\,dQ_{m}\,P\left(I_{m},Q_{m}\right)\text{Tr}\left[\rho^{2}\left(I_{m},Q_{m}\right)\right].
\end{equation}

The relative reduction in purity due to this marginalization procedure is

\begin{equation}
r=1-\frac{\gamma_{\Xi_{\text{Opt}}}}{\gamma_{\text{Avg}}}
\end{equation}

At zero measurement strength, $r$ was calculated to be less than $2\%$, since
no rotation is necessary. For the highest measurement strength sampled
for which fringes still exist, $\Lambda=1.3$, we calculate $r=6\%$, which we judge to be acceptable given the increased SNR on the values of discord. We report the discord of $\rho^{M}_{\Xi_{\text{Opt}}}\left(I_{m}\right)$ in Fig.~\ref{DISCORD}.

As discord is a relatively small effect, we wish to be certain the 
value we extract is a conservative estimate. To obtain confidence intervals which classify
the sampling error, we utilize the bootstrap sampling procedure. We
generated $N=2000$ bootstrap datasets, each of which were subjected
to all previous data processing steps, yielding a distribution of
discords (relative to Alice and relative to Bob) associated with each
measurement strength and $I_{m}$. We expect this distribution to
faithfully represent the sampling error, and thus report the upper
and lower 95th percentile of this distribution as the confidence intervals
associated with measured discord.

\renewcommand{\floatpagefraction}{.8}
\renewcommand{\topfraction}{.8}

\clearpage

\bibliographystyle{apsrev4-1}
\bibliography{Mendeley.bib}

\end{document}